\documentclass[preprint,aps,amssymb,superscriptaddress,
nofootinbib]
{revtex4}
\usepackage{amsmath}
\usepackage{relsize}
\usepackage{accents}
\usepackage{color}
\usepackage{graphicx}
\usepackage{hyperref} 
\hypersetup{
        colorlinks=true,
       linkcolor=black,
        citecolor=black,
        filecolor=black,
        urlcolor=black
}
\definecolor{DarkRed}{rgb}{0.545098,0.000000,0.000000}
\definecolor{blue1}{rgb}{0.000000,0.000000,1.000000}
\definecolor{red1}{rgb}{1.000000,0.000000,0.000000}
\definecolor{indianred}{rgb}{0.803922,0.360784,0.360784}

\newcommand{\Red}[1]{{ \color{red1}{#1}}}

\newcommand{\U}[1]{\underline{#1}}
\preprint{\today \hspace{11cm} IMSc/2017/10/08}
\begin{document}
\title{Quadrupolar power radiation by a binary system in de Sitter background} 

\author{Sk Jahanur Hoque}
\email{skjhoque@cmi.ac.in}
\affiliation{Chennai Mathematical Institute, Siruseri,  Chennai-603 103, India.}
\author{Ankit Aggarwal}
\email{aankita@imsc.res.in}
\affiliation{The Institute of Mathematical Sciences, HBNI,\\  CIT Campus, Chennai-600 113, India.}
\begin{abstract} 
Cosmological observations over past couple of decades favor our universe with a tiny 
positive cosmological  constant. Presence of cosmological constant not only imposes 
theoretical   challenges in gravitational wave physics, it has also observational relevance. 
 Inclusion of cosmological constant in linearized theory of gravitational 
waves modifies the power radiated quadrupole formula. There are two types of observations 
which can be impacted by the modified quadrupole formula. One is the orbital decay of an 
inspiraling binary and other is the modification of the waveform at the detector.  Modelling a 
compact binary system in an elliptic orbit on de Sitter background we obtain {energy and 
angular momentum radiation due to emission of gravitational waves}. We also investigate {evolution of orbital parameters under back reaction and} its impact on orbital decay rate. {In the limit to circular 
orbit our result matches to that obtained in ref. \cite{Bonga}. 
}
\end{abstract}


\maketitle

\section{Introduction}
{Two years after the completion of general relativity, in 1918, Einstein derived power radi- 
ated quadrupole formula in Minkowski background.
Einstein's quadrupole formula was the first 
quantitative estimate of the power radiated in the form of gravitational radiation.
It was also shown that to the leading order the emitted power due to gravitational 
radiation is proportional to the square of the third derivative of mass quadrupole moment of the 
source \cite{Einstein}.}
This power loss would cause binary system to shrink slowly. Such a 
secular change 
in the orbital period of Hulse-Taylor binary pulsar was confirmed by observation
to the accuracy of $10^{-3}$,  thereby providing an 
indirect affirmation of gravitational 
waves \cite{TaylorI,TaylorII,Damour}.
Einstein's theory also passed with flying colors in the direct observation of newly opened
gravitational wave astronomy - gravitational wave template predicted by Einstein's theory 
matches with observed signal \cite{GWI,GWII,GWIII,GWIV}. 

All these theoretical frameworks assume a vanishing
cosmological constant. However, by now cosmological observations (e.g. red
shift of type Ia supernovae) have established that our universe favors a positive 
cosmological constant ($\Lambda$). {Inclusion of cosmological constant not only posits 
theoretical challenges, it could also potentially provide an independent estimate of $
\Lambda$ from the current accuracy of observation}. It is natural 
to ask how does the presence of cosmological constant affect orbital decay of a binary system. 
Change in orbital decay also induces a change in orbital phase which is  sensitive to 
current gravitational wave detectors. The aim of this work is to get an estimate of 
quadrupolar power loss by an elliptic binary system in de Sitter background. A priori the 
order of magnitude of corrections over Minkowski background is not clear. Even if it is negligible, 
it needs to be demonstrated as there are many conceptual and technical difficulties in the 
theory of gravitational waves in de Sitter space-time compared to Minkowski space-time. This 
work is a 
part of AA's M.Sc. thesis \cite{Ankit} and  the case of circular orbit has been discussed in 
JH's Ph.D. thesis \cite{Jahanur}.

While it is well recognized that there is no tensorial (which is both local and generally
covariant) definition of stress tensor for gravitational field, it is possible to construct 
meaningful, quasi-local quantities to represent total energy/momentum in
specific contexts. One of the earliest such proposals is by Isaacson tailored for the 
context in which there are two widely separated spatio-temporal  scales. For sources which 
are rapidly varying (relative to the length scale set by cosmological constant), there is an 
identification of gravitational waves as ripple on a  background within the so called 
`short wave approximation'. Let $L_{B}$ denote the length scale
of variation of the background and $\lambda$ the length scale of the ripple with $
\lambda \ll L_{B}$. 
In this
context Isaacson defined an effective gravitational stress tensor for the ripples which is
gauge invariant to leading order in the ratio of the two scales \cite{Isaacson}. In a recent paper 
\cite{DateJH2} co-authored by one of the authors of this paper, it has been discussed how Isaacson prescription can be adapted to 
compute quadrupolar power, radiated by  a rapidly varying compact source in de Sitter 
background. {Quadrupole formula on de Sitter background can also be obtained using 
covariant phase formalism \cite{ABKII, ABKIII, JHIII}. It exploits the phase space structure of the space of linearized 
solutions and defines a gauge invariant and conserved charges corresponding to the isometries 
of de Sitter background.} Compact binary system is  the 
natural test-bed to apply the 
{\em{new quadrupole 
formula}}.

\ifx
Value of 
cosmological constant being tiny, $10^{-29}$ gm/cc or $10^{-52} m^{-2}$
in the geometrized unit with $G=1=c$, one can neglect $\Lambda$ in the 
vicinity of astrophysical sources or near ground-based gravitational wave detectors. Signature of 
$\Lambda$ may be visible over the vast distances of source-free regions in which 
gravitational waves propagate. To appreciate this statement, let us take the example of 
Schwarzschild-de Sitter space-time in static coordinates,
\begin{equation}
ds^{2}=-\Big(1-\frac{2GM}{r}-\frac{\Lambda r^{2}}{3}\Big)~dt^{2} +
\frac{dr^{2}}{\Big(1-\frac{2GM}{r}-\frac{\Lambda r^{2}}{3}\Big)}+r^{2}d\Omega_{2}^{2} 
\end{equation}
Near the source, the mass term dominates while at the large distance repulsive nature of pure 
de Sitter potential comes into play. Hence, neglecting cosmological term near the source we 
assume that the source dynamics is governed by Newtonian potential and far away from the 
source we relate the energy lost due to gravitational radiation to that of quadrupolar energy flux 
in de Sitter background.
\fi

This paper is organized as follows. In section \ref{Prelim}, we recall the 
power radiated quadrupole formula from \cite {DateJH2}. The {\em modified 
quadrupole formula } involves  mass quadrupole moment as well as pressure 
quadrupole moment of the source. As for non-relativistic (weakly stressed) 
compact source we can neglect the pressure term, in section 
\ref{MassMoment}, we discuss the derivation of mass quadrupole moment of a point particle in 
de Sitter background.  In section \ref{SourceModel}, we spell out the 
assumptions made to model an elliptic binary in de Sitter background. Given 
these assumptions, the quadrupolar energy flux radiated by an elliptic binary 
system is presented in section \ref{BinaryRadiation}, while section \ref{AngularMom} contains a discussion of angular momentum loss .  In section \ref{Evolution}, we discuss decay of orbital parameters due to energy and angular momentum loss. This is a new result. The final 
section \ref{Summary} concludes with a summary and discussion. Some of the technical details 
are given in an appendix. We set $c=1$ throughout. 
%
%
\section{Preliminaries}\label{Prelim}
Cosmological observations over the decades have indicated that our universe is undergoing 
an accelerated expansion, which is most simply {{modelled}} by a positive cosmological 
constant. In this light, the weak gravitational field should be re-looked at as ripples on the de Sitter background.
To compute quadrupolar power radiation due to  a elliptic binary 
system in de Sitter background we follow the framework developed in \cite{DateJH2}.
In this section we review the work and give relevant expressions.
\begin{figure}[h]
\centering
\begin{center}
\includegraphics[width=0.7\textwidth]{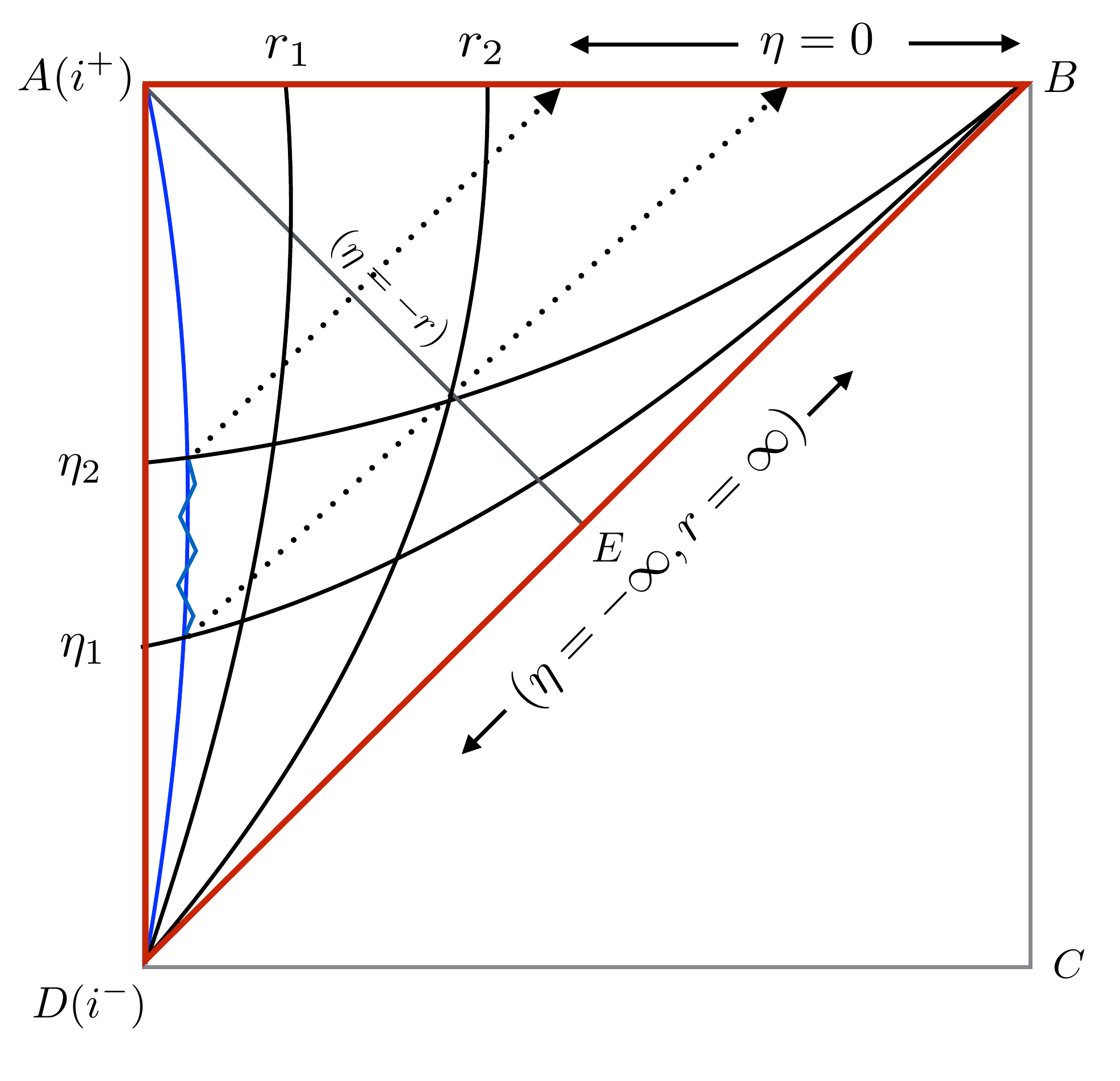}
\caption{The full square is the Penrose diagram of de Sitter space-time
with generic point representing a 2-sphere.  The blue line denotes trajectory of spatially 
compact source in de Sitter background.  ABD part is the the future  Poincar\'e patch,  covered by the 
conformal chart $(\eta, r, \theta, \phi)$. The line AB denotes the {\em future null infinity}, $
\mathcal{J}^+$ while the line AE denotes the {\em cosmological horizon}, $\mathcal{H}^+$. 
Two constant $\eta$ space-like hypersurfaces are shown with $\eta_2 > \eta_1$. The two
constant $r$, time-like hypersurfaces have $r_2 > r_1$.  The two dotted
lines denote the out-going null rays emanating
from $\eta = \eta_1, \eta_2$ on the world line  through the
source. 
}
\label{PoincarePatch}
\end{center}
\end{figure}

The de Sitter space-time defined as the hyperboloid in five dimensional Minkowski space-
time, has a conformal chart  of coordinates ($\eta,r,\theta,\phi$), as shown in figure \ref{PoincarePatch}.  To be 
definite, let us take the world-tube of the spatially compact source to be around the line AD for
all times. In this case source world-tube has future and past time-like infinity, denoted 
by $i^{\pm}$. Causal future of the  compact source is only the future
Poincar\'e patch (ABD) of full de Sitter. No observer whose world-line is confined to the 
past Poincar\'e patch (BCD) can detect the radiation emitted by the gravitating source.
%
Therefore, to study  gravitational radiation due to compact sources, it is sufficient to 
restrict oneself to future Poincar\'e Patch rather than full de Sitter
space-time \cite{DateJHI,ABKII}. There are two natural coordinate charts for the future Poincar\'e patch, i.e., a
conformal chart: $ (\eta,x^{i})$ and a cosmological chart: $(t,x^{i})$. 
In the conformal chart ($\eta, x^i)$ the background de Sitter metric takes the 
form,
\begin{eqnarray}
ds^2  = ~a^{2}(\eta)~ \bigg[ - d\eta^2 + \sum_i (dx^i)^2
\bigg]\ ,~~a(\eta)=-(H\eta)^{-1}~,~ \eta \in ( - \infty , 0 )~,~ H := 
\sqrt{\frac{\Lambda}{3}} . \label{ConformallyFlat}
\end{eqnarray}
While conformal coordinates are convenient in detailed 
calculations of gravitational perturbations, 
they are not suitable for taking the limit $\Lambda \rightarrow 0$ \cite{ABKIII}.
To take that limit we have to use proper time $t$, which is related to conformal time $\eta$ via 
$\eta := - H^{-1}e^{-Ht}$. In these coordinates ($t,x^{i}$), line element becomes,
\begin{eqnarray}   \label{PlanarMetric}
ds^2  =  - dt^2 + e^{2Ht}\sum_{i = 2}^4 (dx^i)^2 
\end{eqnarray}

Poincar\'e patch has seven dimensional symmetry group - 3 spatial rotations, 3 spatial 
translations and 1 time translation. For computing power radiated quadrupole formula in de 
Sitter background, it suffices to focus on the time translational Killing field. In order to find the 
Killing vector field of time translations we let $t\rightarrow t+\delta t$ in the above metric. This 
leads to
\begin{equation}
ds^2  \rightarrow  - dt^2 + e^{2Ht}(1+2H~\delta t)\sum_{i = 2}^4 (dx^i)^2 
\end{equation}
To make the metric invariant under this time translation, the spatial coordinates, $x^{i}$, 
must be transformed as  $x^{i}\rightarrow x^{i}-H x^{i}\delta t$. In general, a Killing vector $
\xi^{\alpha}
$ is an infinitesimal generator of isometry, i.e., for $x^{\alpha}\rightarrow x^{\alpha}+\epsilon 
\xi^{\alpha}$, the metric remains invariant. Identifying $\epsilon=\delta t$, the Killing vector
which generates {time translation in the cosmological coordinate system is 
$T^{\mu}=(1,-Hx^{i})$.
We will work with this time-translational Killing field to compute quadrupolar radiation.

We briefly recall the derivation of quadrupole formula in de Sitter background from 
\cite{DateJH2}.
Under the `short wavelength approximation' for rapidly varying compact source, we visualize 
gravitational waves as high frequency ripple over a slowly varying background space-time. Let 
$L_{B}$ be the length scale variation of background and $\lambda$ be the length scale of 
ripple. In the present context of de Sitter background, $L_{B}$ length scale is set by 
the cosmological constant, $L_{B}\sim\frac{1} {\sqrt{\Lambda}}$.  To maintain a clear cut 
separation between background and ripple for all time, we demand that $\lambda \ll L_{B}$.
In this context, it is possible to define Isaacson effective gravitational stress tensor, 
$t_{\mu\nu}$ for the ripples .
To obtain Isaacson stress tensor, one begins with an expansion of the form
$g_{\mu\nu} = \bar{g}_{\mu\nu} + \epsilon h_{\mu\nu}$ and writes the
Einstein equation  in source free region as,
\begin{eqnarray} \label{ExpandedEqn}
R_{\mu\nu}(\bar{g} + \epsilon h) & = & \Lambda (\bar{g}_{\mu\nu} +
\epsilon h_{\mu\nu})  \nonumber \\
%
\therefore R^{(0)}_{\mu\nu}(\bar{g}) + \epsilon
R^{(1)}_{\mu\nu}(\bar{g}, h) + \epsilon^2 R^{(2)}_{\mu\nu}(\bar{g}, h) 
 & = & \Lambda (\bar{g}_{\mu\nu} + \epsilon h_{\mu\nu}) 
\end{eqnarray}
Introduce an averaging over an intermediate scale $\ell$, $\lambda \ll
\ell \ll L_{B}$, which satisfies the properties: (i) average of odd powers of
$h$ vanishes and (ii) average of space-time divergence of tensors is
sub-leading to the ratio of two length scales, $\lambda/L_{B}$
\cite{MTW,Stein}. Taking the average of the above equation gives,
\begin{equation} \label{BackgroundEqn0}
\langle R^{(0)}_{\mu\nu} \rangle + \epsilon^2\langle R^{(2)}_{\mu\nu}
\rangle = \Lambda\bar{g}_{\mu\nu} .
\end{equation}
Notice that $R^{(2)}$, which is quadratic in $h$, {\em can} have
$L_{B}$ - scale variations and hence non-zero average. Thus it incorporates 
back reaction of ripple on the background and modifies the background equation 
as,
\begin{eqnarray}
8\pi t_{\mu\nu} & = & \bar{R}_{\mu\nu} -
\frac{1}{2}\bar{g}_{\mu\nu}\bar{R} + \Lambda \bar{g}_{\mu\nu}
%
\end{eqnarray}
This can be appropriately termed as `coarse-grained'  form of Einstein equation in source 
free region, where 
\begin{eqnarray}
t_{\mu\nu}(\bar{g}, h) & := & ~- \frac{\epsilon^2}{8\pi}\left[ \langle
R^{(2)}_{\mu\nu}\rangle -
\frac{1}{2}\bar{g}_{\mu\nu}\bar{g}^{\alpha\beta}\langle
R^{(2)}_{\alpha\beta}\rangle \right] \label{RippleTensorDef}
\end{eqnarray}
is the {\em{effective stress-energy tensor}} of ripple. It should be noted that the stress 
tensor is symmetric and conserved with respect to background. 
Given a symmetric, conserved stress tensor and time-translational Killing 
vector of de Sitter  background, one can construct conserved current $J^{\mu}_{T}:=-t^{\mu}
_{\nu}~ T ^{\nu}$. From the conservation equation it follows,
\begin{equation} \label{GeneralConservationEqn}
0 ~ = ~ \int_{\mathcal
V}d^4x\sqrt{\bar{g}}\bar{\nabla}_{\mu}J_{\xi}^{\mu} = \int_{\mathcal
V}d^4x\partial_{\mu}(\sqrt{\bar{g}}J_{\xi}^{\mu}) = \int_{\partial\cal
V}d\sigma_{\mu}J_{\xi}^{\mu} , 
\end{equation}
%
%
For linearized retarded solution in de Sitter background, computing the {\em energy flux}
($\int_{\Sigma}d\sigma_{\mu}J^{\mu}_{T}$) integral across killing hypersurfaces, we obtain 
the power radiated quadrupole formula \cite{DateJH2},
\begin{align}\label{Power}
\mathcal{P} 
 ~ = ~ \frac{G}{8\pi}\int_{S^2}d^2s
\langle\mathcal{Q}_{ij}^{tt}\mathcal{Q}^{ij}_{tt}\rangle  
\end{align}
where,
 $\mathcal{Q}_{ij} :=  \left[\mathcal{L}_{T}^{3}{Q}_{ij} + 3H
\mathcal{L}_{T}^{2}{Q}_{ij} +
2H^2\mathcal{L}_{T}{Q}_{ij} + H\mathcal{L}_{T}^{2}{\bar{Q}}_{ij} + 
3H^2\mathcal{L}_{T}{\bar{Q}}_{ij} +
2H^3\bar{Q}_{ij}\right](t_{ret})
$
and $\mathcal{L}_{T}$ denotes Lie derivative with respect to time 
translational  
Killing vector. We would like to express this quantity in proper time 
coordinate $t$, of  matter source. As moments are function of $t_{ret}$, using $\mathcal{L}_{T}:=
\partial_{t}-2H$ on $Q_{ij} $ and $\bar{Q}_{ij}$,  we can express $\mathcal{Q}_{ij}$ as, 
%
\begin{align} \label{RadiationField}
\mathcal{Q}_{ij}=\left[\partial_{t}^{3}{Q}_{ij} - 3H\partial_{t}^{2}{Q}_{ij} +
2H^2\partial_{t}{Q}_{ij} + H\partial_{t}^{2}{\bar{Q}}_{ij} - H^2\partial_{t}
{\bar{Q}}_{ij} 
\right](t_{ret}).
\end{align}
In these expressions, $Q_{ij}$ and $\bar{Q}_{ij}$ are mass quadrupole 
moment and pressure quadrupole moment 
respectively (see below), while $t_{ret}$ is defined as 
 $\eta-r=-\frac{1}{H}~e^{-Ht_{ret}}$. 
 The label {\em {tt}} denotes algebraically projected  part  of tensor field and 
 is defined as 
$\mathcal{Q}_{ij}^{tt} :=\Lambda_{ij}^{kl} ~\mathcal{Q}_{kl}$. 
 $\Lambda_{ij}^{kl}$ is the algebraic projection operator which projects 
 spatial 
components of a tensor to a plane orthogonal to radial direction and makes it traceless. The 
projection operator is defined as 
$\Lambda_{ij}^{kl}:=
\frac{1}{2}(P_{i}^{k}P_{j}^{l}+P_{i}^{l}P_{j}^{k}-P_{ij} P^{kl}), \mbox{with}~P_{i}^{j} =
\delta_{i}^{j}-\hat{x}_{i}\hat{x}^{j}
$.
\ifx
\Red{
$\Lambda$ projected tensor fields satisfy spatial transversality and tracelessness (TT)
\footnote{Any symmetric rank-2 tensor field can be decomposed as \cite{Gravity},
\begin{equation} \nonumber
A_{ij}=\frac{1}{3}\delta_{ij}\delta^{kl}A_{kl}+(\partial_{i}\partial_{j}-\frac{1}{3}\delta_{ij}
\nabla^{2})B+\partial_{i}B_{j}^{T}+\partial_{j}B_{i}^{T}+A_{ij}^{TT}. 
\end{equation}
$A_{ij}^{TT} $ refers to the transverse-traceless part of the field, so that  $\partial^{i}A_{ij}
^{TT}=0=\delta^{ij} A_{ij}^{TT}$.
} 
condition to the leading order in $r^{-1}$
(for a detailed analysis between {\em tt} vs. TT in the context of asymptotically flat space-time, 
see \cite{ABI,ABII}). Throughout the paper we use projected {\em tt } field. Interested 
readers can take a look at \cite{Bonga} for  derivation of power radiation by a circular  
binary system using TT part of  $\mathcal{Q}_{ij}$. Our main result of the 
paper considers an  elliptic orbit  and we will see in section \ref{BinaryRadiation}, in the limit 
of circular orbit it reduces to the result of \cite{Bonga} (see eq. (25) of the reference). 
}
\fi

Let us take a quick detour to the definition of moments in de Sitter background. To 
maintain coordinate invariance and moment integrals to be well-defined, the moment variables must  
be  coordinate scalars. The natural choice 
is to write moment variables in terms of tetrad components. The conformal form of de Sitter 
metric in eq. \eqref{ConformallyFlat}
suggests a natural choice \cite{DateJHI} , 
\begin{equation}\label{tetrad}
f^{\alpha}_{~\U{0}} ~ := ~ - H\eta~(1, \vec{0}) ~~,~~ f^{\alpha}_{~\U{m}}
~ := ~  - H\eta\ \delta^{\alpha}_{~\U{m}} \hspace{0.7cm}
\Longleftrightarrow \hspace{0.7cm} f^{\alpha}_{~\U{a}} := - H\eta
\delta^{\alpha}_{~\U{a}} 
\end{equation}
The corresponding components of the stress tensor are given by,
%
%
$
\rho :=P_{\underline{00}} =
T_{\alpha\beta}f^{\alpha}_{~\underline{0}}f^{\beta}_{~\underline{0}} ~ = ~
H^2 \eta^2 T_{00} \delta^0_{~\underline{0}}\delta^0_{~\underline{0}}  ,  P_{\underline{ij}} 
~ :=
~ T_{\alpha\beta}f^{\alpha}_{~\underline{i}}f^{\beta}_{~\underline{j}} ~ = ~ H^2\eta^2
T_{ij}\delta^i_{~\underline{i}}\delta^j_{~\underline{j}} \ ;
%
\pi := P_{\underline{ij}}\delta^{\underline{ij}} \ .
%
$
The quadrupole moments of the two rotational scalars, $\rho, \pi$, are defined  by integrating over the source distribution at $t = const.$ hypersurface,
\begin{eqnarray} \label{MassMom}
{Q}^{{i}{j}}(t) & := & \int_{Source(t)} d^3 {x} \ a^{3}(t)
\rho(t,\vec{x}) ~\bar{x}^{{i}} \bar{x}^{{j}} ~~,
\\
 \bar{Q}^{{i} {j}}(t) & := & \int_{Source(t)} d^3 {x} \  a^{3}(t)\pi(t,\vec{x}) ~\bar{x}
 ^{{i}} \bar{x}^{{j}} ~.~
\end{eqnarray}
The determinant of the induced metric on $t=const.$ hypersurfaces is $a^{3}(t)$.
The tetrad components of the moment variable are given by, 
$\bar{x}^{{i}} := f^{\U{i}}_{\alpha} x^{\alpha}
= - (\eta H)^{-1}\delta^{{i}}_{~j}x^{j} = a(t)x^{{i}}$. 
It should be noted that the tetrad components measure the physical distance 
and $d^{3}x~a^{3}(t)=d^{3}\bar{x}$.  Thus, effectively,  moment variable is 
computed in a tetrad frame attached with source.
%
%
\section{ Mass quadrupole moment of a point particle in de Sitter}
\label{MassMoment}
To proceed let us investigate quadrupole moment of a point particle of mass 
$m$ in de Sitter background. For a compact, non-relativistic  source, 
we can neglect pressure with respect to energy density  and  it is sufficient to compute 
only mass  quadrupole moment. 
Let a point particle of mass $m$ is moving on a worldline $\gamma$ in a 
curved space-time with metric $g_{\alpha\beta}$. Action functional of the particle is given by \cite{Gravity},
\begin{equation}
S_{p}=-m \int _{\gamma} \sqrt{-g_{\mu\nu}~\dot {x} ^{\mu}\dot {x} 
^{\nu}} ~ d\sigma
\end{equation}
where $\sigma$ is an arbitrary parameter along worldline of particle (which can be taken 
as proper time for convenience). {We assume that the $m$ is small, so that perturbation 
$h_{\mu\nu}$ created by the particle can also be considered to be small.
Now, as usual, decompose $g_{\mu\nu}=\bar{g}_{\mu\nu}+h_{\mu\nu}$. 
 The background metric $\bar{g}_{\mu\nu}$ is independent of $m$ while the  perturbation 
$h_{\mu\nu}$ contains dependence of $m$.
To the leading order in $m$, particle's energy-momentum tensor in background space-time is 
given by
}
\cite{Poisson, Straumann},
\begin{eqnarray}\label{source}
T^{\mu\nu}=\frac{m}{\sqrt{-\bar g}} \int_{\gamma} u^{\mu} u^{\nu} \delta^{4}(x-x_{p})~d
\tau
\end{eqnarray}
where $u^{\mu}$ denotes particle's four velocity in background space-time and $x_{p}$ is 
the position of the particle. In conformal chart ($\eta,x^{i}$) of de Sitter background 
proper time, $d\tau=\sqrt{\bar{g}_{\mu\nu}~dx^{\mu}dx^{\nu}}=a \sqrt{\eta_{\mu\nu}
~dx^{\mu}dx^{\nu}}=a~\gamma^{-1}d\eta$, where $\gamma^{-1}=\sqrt{1-(\frac{d\vec{x}}
{d\eta})^{2}}$. For computing mass quadrupole moment, let us concentrate on $T^{00}$,
\begin{align}
T^{00}&=\frac{m}{a^{4}}\int \bigg(\frac{d\eta}{d\tau}\bigg)^{2}~\delta^{4}\big(x-x_{p}
(\eta)
\big)~\frac{d\tau} {d\eta}~ d\eta\\
&=\frac{m}{a^{5}} \gamma ~\delta^{3}\big(\vec{x}-\vec{x}_{p}(\eta)
\big)\\
\therefore T^{00} &\approx \frac{m}{a^{5}}  ~\delta^{3}\big(\vec{x}-\vec{x}_{p}(\eta)
\big)
\end{align}
In the last line we have dropped the Lorentz factor $\gamma$ assuming the source is non-
relativistic.
As the tetrad frame (\ref{tetrad}) is defined in ($\eta,x^{i}$) coordinates, we  compute 
$\rho$ in conformal 
coordinates, after that we convert it in ($t,x^{i}$) coordinates. Hence,
\begin{align}
\rho:=a^{-2}(\eta) 
T_{00}~ \delta ^{0}_{\U{0}} \delta ^{0}_{\U{0}}=a^{-2}(\eta) ~\frac{m}{a(\eta)} \delta^{3}
\big(\vec{x}-\vec{x}_{p}(\eta)
\big)~\delta ^{0}_{\U{0}} \delta ^{0}_{\U{0}}= \frac{m}{a^{3}(t)} \delta^{3}
\big(\vec{x}-\vec{x}_{p}(t)
\big)~\delta ^{0}_{\U{0}} \delta ^{0}_{\U{0}}
\end{align}
In the intermediate step we have used conformally flat metric in $(\eta,x^{i})$ to lower 
the indices of $T^{00}$.
Plugging this expression in eqn. (\ref{MassMom}), we obtain mass quadrupole moment of a 
point particle in de Sitter background,
\begin{eqnarray}\label{MassMoment1}
{Q}^{{i}{j}}(t) & =&  m~\bar{x}^{{i}}_{p} ~\bar{x}^{{j}}_{p} ~~.
\end{eqnarray}
In the final expression we have suppressed the constant tetrad $\delta ^{0}_{\U{0}}$. It 
should be noted that the mass  quadrupole moment is expressed in terms of tetrad components. 
In terms of coordinates $(\eta,x^{i})$, tetrad components $\bar{x}^{{i}}:=a(\eta)\delta 
^{\U{i}}_{j}x^{j}$, represent physical distance.
%
%
\section{source modelling in de Sitter background}\label{SourceModel}
To model the binary system in de Sitter background, we follow the strategy 
of flat background \cite{Maggiore, Bounanno}. {In Minkowski space-time 
the ultimate 
destination of all $r=constant$ hypersurface is future timelike infinity, $i^{+}$. 
However in de 
Sitter background,  $r=constant$ hypersurface intersects future null infinity, $
\mathcal{J}^+$.
Any two points on $\mathcal{J}^{+}$ are 
physically infinitely separated. Hence, to maintain finite separation between 
source components,  compact source should be within the cosmological 
horizon and converges on $\mathcal{J}^{+}$ exactly at $i^{+}$ (see fig. 
\ref{BinaryFlux}). For 
example, a circular orbit in de Sitter background is {\em{defined}}  by the  $r_{ph}
=constant$  curve.}

These orbits are not necessarily the 
physical ones, nor do they follow the geodesics of background geometry (in 
fact there is no closed geodesic in de Sitter or flat background). For 
a more realistic scenario, one should investigate the motion of a test particle in 
Schwarzschild  de Sitter background.  
%
{In principle, one can define source moment in Schwarzschild de Sitter background and obtain the linearized field in terms of moments using the conservation of stress tensor. In the far zone, this solution is expected to match with that of linearized field solution of de Sitter background. }
%

%
{We visualize orbital decay of binary system as an iterative process. To start with, say, the system is conservative and
we assume that the orbit is pre-assigned to the de Sitter background.
There is no orbital decay then and the orbital parameters remain constant 
forever. Now from linearized perturbation theory, we obtain the gravitational field 
using conservation of stress-energy tensor of matter source with respect to 
{de Sitter} background. }As the gravitational radiation carries energy it causes the orbit 
to shrink. Hence, we equate the power lost by source to that power 
associated with gravitational waves.
\begin{figure}[h]
\centering
\begin{center}
\includegraphics[width=1.05\textwidth]{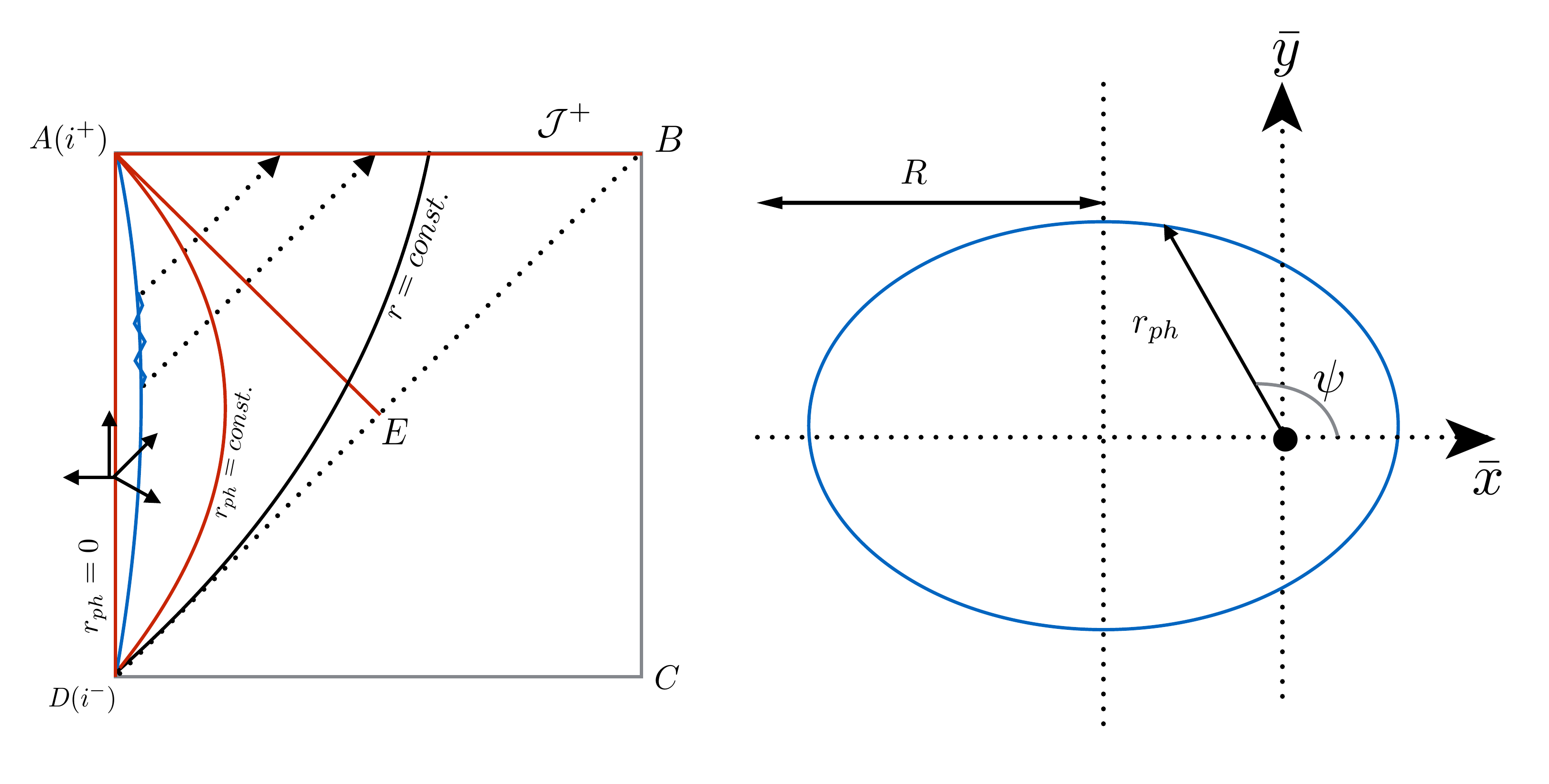}
\caption{In de Sitter background $r=consant.$ 
hypersurfaces intersect $\mathcal{J}^+$. 
 Red lines 
denote $r_{ph}:=ar=consant$ surfaces, which are also Killing surfaces.  AE 
denotes the cosmological horizon ($r_{ph}=H^{-1}$) of the source. The 
spatially compact 
source is well inside the cosmological horizon. Blue line is the worldline of 
source trajectory 
which follows an elliptic orbit in de Sitter background.  A tetrad frame is 
attached to the focus 
of ellipse.}
\label{BinaryFlux}
\end{center}
\end{figure}

We will consider a binary system in an elliptic Keplerian orbit. Therefore in 
the center of mass frame of binary, this system is equivalent to an effective 
one body problem with reduced mass $\mu=\frac{m_{1}m_{2}}{M}$ 
{following an elliptic trajectory, where $M=m_{1}+m_{2}$ is the total 
mass of the system.}
As source moment is defined in tetrad variable, we will attach the tetrad system of conformal 
chart to the center of mass of the system and assign $r_{ph}=0$ to the focus of 
ellipse. We also want to express orbital parameters in terms of physical variables. As 
tetrad component measures physical distance, the  orbital parameters should also be expressed 
in tetrad variable (this automatically incorporates the effect of scale factor in the orbit). In 
terms 
of orbital parameters, eccentricity ($\epsilon$) and semi-major axis ($R$), the equation of orbit 
in {de Sitter background} is {\em{defined}} by,
\begin{align}
r_{ph}=\frac{R(1-\epsilon^{2})}{1+\epsilon \cos \psi}
\end{align}
%
%
{$r_{ph}$ denotes relative physical separation between two components of binary. This 
definition of the orbit is motivated by the fact that any frame observer of de Sitter background 
measures the orbital separation as an ellipse. 

As 
discussed earlier, moments should be  defined in the tetrad frame.} We choose the 
time direction of tetrad frame along the worldline {(measures the proper time $t$)} of source 
and a triad frame $(\bar{x}, \bar{y}, \bar{z})$ is attached to the focus of the ellipse such that 
the orbit is restricted to $(\bar{x}, \bar{y})$ plane and is given by,
\begin{align}
\bar{x}(t)=r_{ph}~ \cos{\psi}~~,~~~
\bar{y}(t)=r_{ph}~ \sin{\psi}~~,~~~
\bar{z}(t)=0
\end{align}
Hence, from eq. \eqref{MassMom} the mass quadrupole moment of the binary system can be 
written in a matrix form ,
\begin{equation}
  {Q}^{ij}=\mu {r_{ph}^{2}}~
  \begin{pmatrix}
   \cos ^{2}\psi & \sin \psi \cos \psi & 0 \\
    \sin \psi \cos \psi & \sin ^{2}\psi  & 0 \\
   0 & 0 & 0
   \end{pmatrix}
\label{MatrixMoment}
\end{equation}
%
%

We restrict our analysis to the adiabatic limit, i.e., the time scale for orbital parameters to change is much longer than orbital time period. 
For example, this is equivalent to saying  $\mid \frac{\Delta E}{E} 
\mid \ll 1$, where $\Delta {E}$ denotes change in energy of the system over one period and $E$ is the total energy of the system. This can be seen easily from the leading term of eq. \eqref{ElBinaryPow}, $\mid \frac{\Delta E}{E} 
\mid = \mid \frac{\langle P \rangle T_{P}}{E} \mid \sim \frac{\mu}{M}\big( \frac{GM}
{R}\big)^{5/2} \ll 1$. At this point, it is worthy to mention that to study the secular change in orbital parameters one needs to investigate the change over a period, only instantaneous change does not help. For a circular orbit, instantaneous power does not have angular dependence (see eq. 
\eqref{InstPow}), hence average over 
one time period is same 
as the unaveraged answer. But for elliptic orbit instantaneous power  has angular 
dependence, hence {\em averaging plays a crucial role.}  
In adiabatic approximation, the orbital parameters, semi-major axis and eccentricity of 
elliptic orbit 
(or equivalently energy and angular momentum) are assumed to be constant 
of motion over one orbital time period. The system spends many periods 
near any point of its phase space trajectory.  As the gravitational radiation 
carries both energy and angular momentum, the binary system undergoes 
secular changes, both in its semi-major axis and eccentricity. For our purpose, 
in the next section,  we concentrate on energy loss and consequent orbital 
decay rate (time derivative of the period).

\section{ Power radiation from inspiraling binary } \label{BinaryRadiation}
%
At first we would like to compute power radiated by an elliptic inspiraling 
binary system due to {\em new quadrupole formula} in eq. \eqref{Power}. Now using $\Lambda$ projection,
$\mathcal{Q}_{ij}^{tt}\mathcal{Q}^{ij}_{tt}:=\Lambda_{~ij}^{kl}\mathcal{Q}
_{kl} \Lambda^{ij}
_{~mn} \mathcal{Q}^{mn}
=\Lambda^{kl}_{~mn} \mathcal{Q}_{kl} \mathcal{Q}^{mn}$, eq. 
(\ref{Power}) can be written as,
\begin{align}\label{ProjectedPower}
\mathcal{P}=\frac{G}{5}\big\langle \mathcal{Q}_{ij}\mathcal{Q}^{ij}-
\frac{1}{3} \mathcal{Q}^{2}\big\rangle,
\end{align}
{where $\mathcal{Q}:= \delta^{ij}\mathcal{Q}_{ij}$}. In deriving this expression we 
have used 
the identity for $\Lambda$ projector,
$\int d^{2}S~ \Lambda^{ij}_{~kl}=\frac{2\pi}{15}~\bigg[11\delta^{i}_{k}\delta^{j}
_{l}-4\delta^{ij}_{kl}+\delta^{i}_{l}\delta^{j}_{k}\bigg]$.
For a weakly stressed non-relativistic system, as for Newtonian fluids, pressure can be 
neglected compared to the energy density, so we can neglect the 
pressure quadrupole moment terms in eq. \eqref{RadiationField}.
%
Hence neglecting the pressure quadrupole moment terms,  power radiated by 
binary system can be expressed as,
\begin{align} 
\mathcal{P}&=\frac{G}{5}\big\langle \mathcal{Q}_{ij}\mathcal{Q}
^{ij}-\frac{1}{3} \mathcal{Q}^{2}\big\rangle\\
& \approx \frac{32G^{4}\mu^{2}M^3}{5  R^5} ~f(\epsilon)
+~\frac{8 H^{2} G^{3} \mu^2 M^2}{ R^2} ~g(\epsilon)
+~ \frac{8 H^{4} G^{2} \mu^{2} M R}{  5} ~ h(\epsilon)
\label{ElBinaryPow}
\end{align}

where $f(\epsilon), g(\epsilon), h(\epsilon)$ are

\begin{align} \label{EnhanceFun1}
f(\epsilon)&=(1-\epsilon^2)^{-7/2}~\bigg(1+\frac{73}{24}\epsilon^{2}+\frac{37}{96}
\epsilon^{4}\bigg),\\
g(\epsilon)&=(1-\epsilon^{2})^{-1/2}~\bigg[\frac{4-\sqrt{1-\epsilon^ 2}}{3}\bigg],\\
h(\epsilon)&=1-\frac{\epsilon^2}{3}.  \label{EnhanceFun2}
\end{align}
Note that each of these functions approaches unity as $\epsilon \rightarrow 0$. 
Schematics of the quadrupolar power computation are given in  appendix 
\ref{Apn1}.  
Order $H$ and $H^{3}$ terms exactly vanish, after taking time average over orbital period. This can also be understood by the following observation. When direction of time is reversed, orbit goes from anticlockwise to clockwise direction and radiated energy does not depend on the orientation of revolution.  
The order $H$ and $H^3$ terms in \eqref{InstPow} have been generated 
from odd number of time derivatives of moment and thus have odd parity 
under time reversal and therefore have to vanish.
For $\Lambda \rightarrow 0$, this formula reduces to the usual formula for 
quadrupolar power radiation by the binary system in an elliptic  orbit  
\cite{Peters}.

To know the order of magnitude, for simplicity let us take a look at power radiated by a 
circular orbit. Taking $\epsilon\rightarrow 0$ in the eq. \eqref{ElBinaryPow} we obtain the 
expression for circular orbit,
\begin{align} \label{CircPow}
\mathcal{P}\approx \frac{32G^{4}\mu^{2} M^{3}}{5~ R^{5}}~
\bigg[1+\frac{5}{4}~\frac{H^{2}R^{3}}{GM}+\frac{1}{4}~\frac{H^{4}R
^{6}}{G^{2}M^{2}}~\bigg]
\end{align}
This expression has the dimensionless expansion parameter
 $\frac{H^{2}R^{3}}{GM}$. As mentioned in \cite{Satya}, 
a compact binary that coalesces after passing through the last stable orbit is a 
powerful source of gravitational waves, we assume $R=r_{LSO}=3 R_{S}$. Take $m_{1}=m_{2}=1 M_{\odot}$ and using Schwarzschild radius of sun, $R_{S}=2GM_{\odot}
=3\times 10 ^{3} meters$,
\begin{align}
\frac{H^{2}R^{3}}{GM}=\frac{2 H^{2}R_{S}^{3}}{R_{S}} \times 27 = H^{2}
R_{S}^{2} \times 54 \approx 10^{-52} \times 10 ^{6}\times 10^{2} \approx 10^{-44}
\end{align}

It is customary and convenient to express \eqref{CircPow} in terms of { \em 
chirp mass} and gravitational wave frequency.  For circular orbit substituting 
$\omega_{s}=\sqrt{\frac{GM}{{R}^{3}}}$. Hence, power loss due to 
gravitational radiation from a circular binary orbit can be expressed as,
\begin{align}
\mathcal{P} &\approx\frac{32 G\mu^{2}R^{4}\omega_{s}^{6}}{5}~\bigg( 
1+\frac{5}{4} \label{OmegaPow} \nonumber
H^{2}\omega_{s}^{-2}+\frac{1}{4}{H^{4}\omega_{s}^{-4}}\bigg)\\
&=\frac{32}{5 G}~\bigg(\frac{G M_{c} ~\omega_{gw}}{2}
\bigg)^{10/3}\bigg[1+5H^{2}\omega_{gw}^{-2}+{4}H^{4}
\omega_{gw}^{-4}
\bigg]
\end{align}
where we introduce the {\em chirp mass} $M_{c}:={\mu}^{3/5} {M}^{2/5}$ 
and $\omega_{gw}=2\omega_{s}$. 
This result matches with that of 
\cite{Bonga} (see eq. (25) of the reference).
{It is also clear from eq. (\ref{OmegaPow}) 
- correction terms are down by $H^{2}/
\omega_{gw}^{2}$, e.g., for LIGO (frequency band  $f\sim 10^{2}-10^{3}$ Hz),  $H^{2}/\omega^{2}_{gw}
\sim  (\lambdabar/L_{B})^{2}\sim 10^{-42} - 10^{-44}$, while for PTA ($f\sim 
10^{-6}-10^{-9}$ Hz), the order of magnitude of leading order correction term  is $ 
10^{-20}-10^{-26}$. Though the correction terms are more significant for low frequency band detectors, they are still negligible.
}
\section{angular momentum radiation}\label{AngularMom}
{It is well known that Isaacson stress-energy tensor does not  suffice to 
capture flux of angular momentum even in Minkowski background
\cite{Gravity}. Nevertheless, one can employ the formalism of covariant 
phase space to get a correct expression for angular momentum flux. We use 
the formalism developed in \cite{ABKIII} to compute angular momentum flux 
radiated by a binary system in de Sitter background. The instantaneous angular momentum radiation in de Sitter background is given by\cite{ABKIII},
\begin{equation}
\dot L_{k}=\frac{G}{4\pi} \ \epsilon_{ik} \  ^{m}\int d^2 S \big(\mathcal{Q}^{ij}
\big) \big(\mathcal{L}_{T}^{2}Q_{jm}+H\mathcal{L}_{T} Q_{jm}+H \mathcal{L}_{T}\bar{Q}
_{jm}+H^{2}\bar{Q}_{jm}\big)^{TT}
\end{equation}
Average angular momentum radiated by an elliptic binary over a orbital time 
period is given by after performing an $S^{2}$ integral, 
\begin{equation} \label{AvgAngMom}
\big\langle \dot L_{k} \big\rangle=\frac{2G}{5} \epsilon_{ik} \  ^{m} 
\Big\langle\big(\mathcal{Q}^{ij}
\big) \big(\mathcal{L}_{T}^{2}Q_{jm}+H\mathcal{L}_{T} Q_{jm}+H \mathcal{L}_{T}\bar{Q}
_{jm}+H^{2}\bar{Q}_{jm}\big)\Big\rangle, 
\end{equation}
where in deriving this expression we have used  local projector to extract the 
TT part. Neglecting the 
pressure quadrupole moments, the expression for  angular momentum 
radiation can be found in 
the same manner as energy radiation. As the binary system is restricted in 
$x-y$ plane, the $z$ component of emitted   
angular momentum flux  (over an orbital period)  is given by,
\begin{align} 
\big\langle\dot L_{z}\big\rangle &= \frac{2G}{5}  \epsilon_{iz} \  ^{m} 
\Big\langle\big(\mathcal{Q}^{ij}
\big) \big(\mathcal{L}_{T}^{2}Q_{jm}+H\mathcal{L}_{T} Q_{jm}+H \mathcal{L}_{T}\bar{Q}
_{jm}+H^{2}\bar{Q}_{jm}\big)\Big\rangle \\
&\approx \frac{32G^{7/2}\mu^{2}M^{5/2}}{5  R^{7/2}} ~F(\epsilon)
+~\frac{8 H^{2} G^{5/2} \mu^2 M^{3/2}}{ R^{1/2}} ~G(\epsilon)
+~ \frac{8 H^{4} G^{3/2} \mu^{2} M^{1/2} R^{5/2}}{  5} ~ H(\epsilon)
\label{AvgAngMom1}
\end{align}
where $F(\epsilon), G(\epsilon), H(\epsilon)$ are

\begin{align}
F(\epsilon)&=\frac{1}{(1-\epsilon^2)^{2}}~\big(1+\frac{7}{8}\epsilon^{2}\big),\\
G(\epsilon)&=(1-\epsilon^{2})^{1/2},\\
H(\epsilon)&=(1-\epsilon^{2})^{1/2}(1+\frac{3}{2}\epsilon^2).  \label{EnhanceFun3}
\end{align}
It should be noted that each of these functions approaches unity as $\epsilon
\rightarrow 0$. 
Further details of the above computation are given in appendix \ref{Apn1}. 
Here also 
order $H$ and $H^3$ do not contribute. For $t \rightarrow -t$, the direction 
of angular 
momentum gets flipped. The order $H$ and $H^3$ terms in angular momentum flux 
computation have been generated from even number of time derivatives of moment and thus 
have even parity under time reversal and therefore have to vanish. From eq. 
\eqref{ElBinaryPow} 
and \eqref{AvgAngMom1} it is evident that for circular case $\langle \dot E \rangle =
\omega_{s} \langle \dot L_{z}\rangle$.

\section{evolution of orbital parameters}\label{Evolution}
The results of previous two sections can be applied to compute secular change in orbital parameters. A priori it is not clear whether the conservative dynamics of binary is governed by Newtonian Potential. We assume that near the source the space-time geometry is Schwarzschild de Sitter. 
In the weak field limit around flat 
background repulsive nature of de Sitter potential comes into play and 
correction terms generated by this potential is also of same order $
\frac{H^{2}r_{ph}^{3}}{GM}$ (it is clear from eq. \eqref{SdS} in appendix 
\ref{Apn2}). But we are analyzing perturbation around de Sitter 
background and we relate field solution in terms of source moments using 
conservation equation on de Sitter background. Therefore, it is necessary to 
investigate weak field limit on pure de Sitter background. A full resolution 
of this problem requires expansion over Schwarzschild de  Sitter metric using 
the technique of black hole perturbation theory which is beyond the scope of 
this work. In appendix \ref{Apn2}, we have argued that in the 
weak field limit on de Sitter background the potential can be still be assumed 
to be Newtonian. The repulsive nature of de Sitter potential can be absorbed in background itself.  Hence the orbital parameters $R, \epsilon
$ are related to $E$ and $L$ through following 
equations,
\begin{eqnarray}
R=-\frac{GM\mu}{2E},\\
\epsilon^{2}=1+\frac{2EL^2}{G^2M^2\mu^3}
\end{eqnarray}
Using equations \eqref{ElBinaryPow}, \eqref{AvgAngMom1} we obtain $\langle \dot R \rangle$ and $\langle\dot\epsilon\rangle$,
\begin{eqnarray} \label{Radius}
\langle {\dot R}\rangle &=&-\bigg(\frac{64}{5}\frac{G^3 \mu M^2 }{R^3} 
f(\epsilon)
+16 H^2 G^2 \mu M g(\epsilon)+\frac{16}{5} H^4 G \mu R^3 h(\epsilon)
\bigg)\\ \nonumber
\langle \dot \epsilon\rangle  = &-&\frac{304}{15}\frac{G^3 \mu M^2}
{R^4} \label{Eccentricity}
\frac{\epsilon}{(1-\epsilon^2)^{5/2}}\Big(1+\frac{121}{304}\epsilon^2\Big) \\
&-&\frac{32}{3} \frac{H^2G^2\mu M}{R} \frac{(1-\epsilon^2)^{1/2}}{\epsilon}\big(1-\sqrt{1-
\epsilon^2}\big) 
+ \frac{44}{15} H^{4}G\mu R^2 \epsilon (1-\epsilon^2)
\end{eqnarray} 
%
%
For $H\rightarrow 0$, these results match with usual flat space  results.
From eq. \eqref{Eccentricity} we see \footnote{For coefficient of $H^2$, $\lim_{\epsilon \to 
0} \frac{(1-\epsilon^2)^{1/2}}{\epsilon}\big(1-\sqrt{1-
\epsilon^2}\big) =0$. } that for $\epsilon\rightarrow 0$, $
\langle \dot\epsilon \rangle=0$. Therefore, circular orbit remains circular. 
It should also be noted that the contribution from $H^4$ term is positive but the smallness of 
$H^2R^3/GM$ ensures that $H^2$ term dominates making the overall contribution negative. 
For an elliptic orbit $0<\epsilon<1$, eq. \eqref{Eccentricity}
gives $\langle \dot\epsilon \rangle<0$. Therefore, elliptic orbit becomes 
more and more circular due to emission of gravitational waves. To get the 
time to coalescence for circular orbit one needs to integrate eq. 
\eqref{Radius}. We do not give the full expression to avoid the cluttering, in this case also   
leading order correction term is of the order $H^2 R^3/GM$.

Now, let us investigate an observable parameter $\dot{P_{b}}$ (time 
derivative of orbital period) for elliptic orbit. Orbital period $P_{b}$ is related 
to orbital energy as $P_{b}=constant
\times (-E)^{3/2}$. Hence,
\begin{align}
\frac{\dot{P}_{b}}{P_{b}}&=-\frac{3}{2}~\frac{\dot{E}}{E}
\end{align}
Assuming that the binary system is loosing energy entirely due to quadrupolar radiation, 
substituting eq. \eqref{ElBinaryPow} in $\dot{E}$, we find
\begin{equation} \label{DecayRate}
{\dot{P}_{b}}=-\frac{192\pi}{5}~{G^{5/3}\mu M^{2/3}}~
\bigg(\frac{P_{b}}{2\pi}\bigg)^{-5/3}
f(\epsilon)\times \bigg[ 1+{\frac{5}{4}} H^{2} \bigg(\frac{P_{b}}
{2\pi}\bigg)^2 ~
\frac{g(\epsilon)}{f(\epsilon)}+{\frac{1}{4}} H^{4}\bigg(\frac{P_{b}}
{2\pi}\bigg)^4 ~\frac{h(\epsilon)}{f(\epsilon)} \bigg],
\end{equation}
where the average over an orbital period is understood. In the intermediate 
step we have 
also used $E=-\frac{G\mu M}{2R}$ and $P_{b}=2\pi\sqrt{\frac{R^{3}}{GM}}
$. For Hulse-Taylor 
Pulsar the values of relevant parameters are: $m_{1}=1.4414M_{\odot}, 
m_{2}
=1.3867M_{\odot}, \epsilon=0.617338, P_{b}=2.790698\times10^{4}s$ \cite{Satya}. 
Hence, for Hulse-Taylor 
binary first order correction for orbital decay rate,
\begin{equation}
\frac{5}{4} H^{2} \bigg(\frac{P_{b}}{2\pi}\bigg)^2 ~\frac{g(\epsilon)}
{f(\epsilon)}
=\frac{5}{4}\times \frac{10^{-35}}{3}\times \bigg(\frac{P_{b}}{2\pi}\bigg)^2 ~
\frac{g(\epsilon)}{f(\epsilon)}\sim 10^{-29}
\end{equation}
Current accuracy level of the observation of  orbital decay rate of Hulse-Taylor pulsar is at $10^{-3}$.
Hence correction terms due to $\Lambda$ are {utterly negligible} in the current observational  context.
%
}
\section{{Comparision between }TT vs \lowercase{tt}}
In this section, we compare our result with that of obtained in a 
recent paper \cite{Bonga}. 
The procedure to obtain power emitted by elliptic orbit in 
our work is 
completely different from the ref. \cite{Bonga}. In our paper we use local algebraic tt projector to obtain the power 
radiation while  ref. \cite{Bonga} relies on extracting transverse traceless (TT) part 
of source quqdrupole moment. For clarity, let us recall that
any symmetric rank-2 tensor field can be decomposed as \cite{Gravity},
\begin{equation} \nonumber
Q_{ij}=\frac{1}{3}\delta_{ij}\delta^{kl}Q_{kl}+(\partial_{i}\partial_{j}-\frac{1}{3}\delta_{ij}
\nabla^{2})B+\partial_{i}B_{j}^{T}+\partial_{j}B_{i}^{T}+Q_{ij}^{TT}. 
\end{equation}
$Q_{ij}^{TT} $ refers to the transverse-traceless part of the field, so that  $\partial^{i}Q_{ij}
^{TT}=0=\delta^{ij} Q_{ij}^{TT}$.
Algebraically projected {\em{tt}} fields are is defined as $Q_{ij}^{tt}:=\Lambda_{ij}^{kl}Q_{kl}$, with
$\Lambda_{ij}^{kl}:=
\frac{1}{2}(P_{i}^{k}P_{j}^{l}+P_{i}^{l}P_{j}^{k}-P_{ij} P^{kl})  \ \mbox{and}~P_{i}^{j} =
\delta_{i}^{j}-\hat{x}_{i}\hat{x}^{j}
$.
$\Lambda$ projected tensor fields satisfy spatial transversality and tracelessness (TT)
condition to the leading order in $r^{-1}$ \cite{DateJH2}.
For a detailed analysis between {\em tt} vs. TT in the context of asymptotically flat space-time, 
see \cite{ABI,ABII}. 
A priori these two notions are distinct and inequivalent. Although these two notions match at future null infinity for Minkowski space-time, 
 they are quite different in de Sitter space-time.  The tt-projection is well tailored to
the $1/r$ expansion commonly used for asymptotically flat spacetimes.
As the global structure of de Sitter spacetime is different from Minkowski spacetime, 
expansion in powers of $1/r$ is not a useful tool to analyse asymptotically de Sitter spacetimes. 
In particular, the tt-projection is not a valid operation to extract the transverse-traceless part of a rank-2 tensor on the full $\mathcal{J}^+$. The TT-tensor is the correct notion of transverse traceless tensors. However, if one restricts oneself to large radial distances away from the source, one may expect that the tt-projection also gives useful answers.
 We stick to tt projection throughout our paper 
 and emphasize that in the context of power radiated by a circular binary system in de Sitter 
 background the answer exactly matches with that of computation done in ref \cite{Bonga} 
 using TT (see eq. (25) of the reference). Though explicit expression of $Q_{ij}^{TT}$ and $Q_{ij}^{tt}$ are quite different (see eq. (20) of ref. \cite {Bonga}) even in $\mathcal{J}^+$ of de Sitter space-time, after integration over two-sphere in energy flux formula the resulting expression for power is the same. This observation is new and  indicates that for energy 
 flux computation  by a circular binary, TT  vs {\em{tt}} {\em does not matter in de Sitter background}. 
\footnote{It is mentioned in \cite{ABI, ABII} for the energy flux computation in 
asymptotically flat space-time, identification of {\em tt} vs TT does not 
matter.}
{We also compare our angular momentum flux expression to the leading 
order expression obtained   in 
\cite{Bonga} (see eq. (26) of the reference) for circular orbit. To the leading order in $H^{0}$ it matches 
exactly. We expect the higher order correction terms in $H$ will also match as the tt operation 
generates an overall factor.}
\section{ summary and discussion} \label{Summary}

In this short paper, we obtain quadrupolar energy and angular momentum loss due to gravitational radiation for a generic elliptic binary system in de Sitter background. We also discuss the decay of eccentricity and semi-major axis due to gravitational radiation.  As noted earlier, the {\em modified quadrupole formula} can impact
both direct and indirect observations of gravitational waves. While indirect observations track orbital decay rate of the binary system, direct observation in gravitational wave astronomy is sensitive to changes in the orbital phase. This paper focuses on the former and concludes that impact of $\Lambda$ in orbital decay is negligible in the context of current
accuracy of observations.
%
%
From dimensional analysis one may argue that correction terms due to $\Lambda$ should be $
\sqrt{\Lambda}\times lengthscale$. This correction may be relevant over cosmological distances, e.g. mega-parsec. There are two natural length scales - one is observational distance and another is source dimension. It should be noted that though linearized field expression depends on observational length scale, energy and angular momentum flux are independent of observational length scale. 
Hence, only available length scale is orbital length scale which enters into the expression via the definition of source quadrupole moments. A typical compact object has orbital extension to the order of $10^{6} ~m$. Hence, this crude analysis also suggests that the correction terms 
should be of the order of $\sqrt{\Lambda}\times orbital~lengthscale\sim 10^{-26}\times 10^{6}
=10^{-20}$.
In our case, the correction terms are even smaller as order 
$H(\sqrt{\Lambda/3}) $ term drops out and leading correction term is of 
the order $H^{2}$. 

{In this paper we also emphasize that for power radiated by a circular binary in de Sitter 
background, identification of TT vs tt does not matter. Why these two completely distinct  
notions give the same answer for energy flux of circular orbit  in de Sitter space-time is still 
needed to be understood clearly. Whether 
this result generalizes to ellipitic orbit case or  for other sources is still needed to be 
investigated. }
\ifx
\begin{figure}[h]
\centering
\begin{center}
\includegraphics[width=0.7\textwidth]{sharpProp.pdf}
\caption{Blue line denotes trajectory of source in de Sitter background while the red lines are 
Killing surfaces. The source is active between the time interval $t_{1}$ to $t_{2}$. As the 
energy propagation is sharp, there is no energy flux across the out-going null surfaces. $t=
\bar{t}_{1}$ and $t=\bar{t}_{2}$ denote two out-going null rays. $\bar{t}$ is defined via 
retarded time $\eta-r=-\frac{1}{H}e^{-H\bar{t}}$.  $\tau$ is Killing parameter. It 
is shown in \cite{DateJH2} that between two outgoing null rays, the Killing 
interval $d\tau$ is 
same for different Killing surfaces, it also equals to $dt$. Therefore, energy lost by 
source in the time interval $dt$ is registered on $\mathcal{H}^+$ or $\mathcal{J}^+$ along 
the out-going null direction. }
\label{SharpProp}
\end{center}
\end{figure}

A question regarding the validity of eq. \eqref{DecayRate} (where we related rate of energy lost by binary to that of quadrupolar power of gravitational waves) arises due to non 
availability of Bondi-type mass loss formula relating to flux in de Sitter background. Recall from 
\cite{DateJH2}, though linearized field in de Sitter background has a `tail' term, in the energy propagation tail term does not contribute. This is because energy depends on the time 
derivative of field, not only the field. In the process of taking the derivative, tail term cancels out exactly, leaving the sharp propagation. Hence, flux across out-going null hypersurface vanishes.
Along the Killing trajectory $d\eta/d\tau=-H\eta, dr/d\tau=-Hr$; therefore $\eta-r=(\eta_{*}-
r_{*}) e^{-H\tau}=(-\frac{1}{H} e^{-H\bar{t_{*}}})~e^{-H\tau}$. Again from the definition of $
\bar{t}$, $\eta-r=-\frac{1}{H}e^{-H\bar{t}}$. Hence $\tau=\bar{t}-\bar{t}_{*} \implies d
\tau=d\bar{t}$. As $\bar{t}$ is constant along null ray we can compute $d\bar{t}$ on source 
trajectory which is $dt$. Now if we assume that the source is losing energy only via gravitational 
radiation, energy  lost by the source in the time interval $dt$ is exactly 
the energy flux across  gravitational wave at  the portion $\mathcal{H}^{+}$ or $\mathcal{J}^+
$ bounded by two out-going null lines (as shown in fig. \ref{SharpProp}).
\fi
\acknowledgements We would like to thank Ghanashyam Date for numerous discussion sessions and improvement of the initial draft. JH also acknowledges fruitful discussions with K G Arun. JH thanks to Amitabh Virmani for a careful reading of the draft. The work was initiated and partially done at 
The  Institute of Mathematical Sciences, HBNI, and JH thanks IMSc for hospitality during the preparation of the manuscript.  The work of JH is supported in 
part by the DST-Max Planck Partner Group ``Quantum Black Holes'' between 
AEI, Potsdam and CMI, Chennai.

\appendix 
\section{Energy and angular momentum  radiated by an elliptic binary}  \label{Apn1}
%
As computation of radiated {energy and angular momentum} need derivative of moments, let us give the expressions for 
derivative of moments. Taking the quadrupolar mass moment tensor \eqref{MatrixMoment} for 
elliptic orbit, derivatives of different components are given by,
\begin{align}
\dot{Q}_{11}&=\dfrac{-\alpha}{1+\epsilon\cos\psi}\sin{2\psi}, \\
\dot{Q}_{12}&=\dfrac{\alpha}{1+\epsilon\cos\psi}(\epsilon\cos\psi+\cos{2\psi}),\\
\dot{Q}_{22}&=\dfrac{2\alpha}{1+\epsilon\cos\psi}(\epsilon\sin{\psi}+\sin{\psi}
\cos{\psi}),~~~~~~~~~~~~~~~~
\mbox{ with $ \alpha=\mu\sqrt{GMR(1-\epsilon^2)}.$}
\end{align}
\begin{align}
\ddot{Q}_{11}&=-\beta[2\cos{2\psi} (1+\epsilon\cos\psi)+e\sin{2\psi}\sin\psi)],\\
\ddot{Q}_{12}&=-2\beta[\sin{2\psi} +\epsilon\sin{\psi}(1+\cos^2\psi)],\\
\ddot{Q}_{22}&=2\beta[\cos{2\psi}+\epsilon\cos{\psi}(1+\cos^2\psi)+\epsilon^2],~~~~~~~~~~~~~~~~~~~~~~~~~
\mbox{ with $\beta=\frac{\mu GM}{R(1-\epsilon^2)}.$}
\end{align}
%
%
\begin{align}
\dddot{Q}_{11}&=\gamma(1+\epsilon\cos\psi)^2[2\sin{2\psi}+3\epsilon\, \sin\psi \,
\cos^2\psi],\\
\dddot{Q}_{12}&=\gamma(1+\epsilon\cos\psi)^2[-2\cos{2\psi}+\epsilon\cos
\psi(1-3\cos^2\psi)],\\
\dddot{Q}_{22}&=-\gamma(1+\epsilon\cos\psi)^2[2\sin{2\psi}+\epsilon\sin
\psi(1+3\cos^2\psi)],~~~~~~
\mbox{ with $\gamma=\frac{2\mu(GM)^{3/2}}{R^{5/2}(1-\epsilon^2)^{5/2}}.$}
\end{align}
%
%
The expression for radiated power is given by \eqref{ProjectedPower},
\begin{align}
\mathcal{P}=\frac{G}{5}\big\langle\mathcal{Q}_{ij}\mathcal{Q}_{ij}-
\frac{1}{3}\mathcal{Q}^{2}\big\rangle
:=\langle P(\psi)\rangle
\end{align}
Now we plug the derivatives of moments in computing unaveraged quantity $P(\psi)$ 
and express our result 
order by order in $H$. 
%
\ifx
Order $H^{0}$ term in power expression,
\begin{align}
\frac{8G^{4}}{15~ c^{5}}~\frac{\mu^{2}M^{3}}{R^{5}(1-\epsilon^2)^{5}}
(1+\epsilon\cos{\psi})^{4}~\big[12(1+\epsilon\cos{\psi})^{2}+\epsilon^{2}\sin^{2}{\psi}
\big]
\end{align}

Order $H$ term,
\begin{align}
\frac{4G^{7/2}}{5~c^{5}}~\frac{\epsilon~ \mu^{2}M^{5/2}}{R^{7/2}(1-\epsilon^{2})^{7/2}}~
\sin{\psi}(1+\epsilon \cos{\psi})^{2}~
[18+13\epsilon^2+40 \epsilon \cos{\psi}+9\epsilon^{2} \cos{2\psi}]
\end{align}

Order $H^2$ term,
\begin{align}
\frac{2G^{3}}{15~c^{5}}~\frac{\mu^2 M^2 }{(1-\epsilon^2)^2~R^2}~
\big[60+100\epsilon^2+36\epsilon^4+\epsilon(180+113\epsilon^2)\cos\psi
+116\epsilon^2\cos2\psi+19\epsilon^3\cos3\psi\big]
\end{align}

Order $H^{3}$ terms,
\begin{align}
-\frac{32 G^{5/2}}{5~c^{5}}~\frac{\epsilon^{2}\mu^2 M^{3/2}}{\sqrt{R(1-\epsilon^{2})}}~\frac{\sin\psi(\epsilon+\cos\psi)}{1+\epsilon\cos\psi}
\end{align}

Order $H^{4}$ terms,
\begin{align}
\frac{4G^{2}}{15~c^{5}}~\frac{\mu^2 M R(1-\epsilon^2)}{(1+\epsilon\cos\psi)^{2}}~\big[6+7\epsilon^{2}+12\epsilon\cos\psi+\epsilon^{2}\cos2\psi\big]
\end{align}
\fi
\begin{align}\label{InstPow}
P(\psi)\approx &\frac{8G^{4}}{15}~\frac{\mu^{2}M^{3}}{R^{5}(1-\epsilon^2)^{5}}
(1+\epsilon\cos{\psi})^{4}~\big[12(1+\epsilon\cos{\psi})^{2}+\epsilon^{2}\sin^{2}{\psi}
\big]\\ \nonumber
+&\frac{4G^{7/2}H}{5}~\frac{\epsilon~ \mu^{2}M^{5/2}}{R^{7/2}(1-\epsilon^{2})^{7/2}}~
\sin{\psi}(1+\epsilon \cos{\psi})^{2}~
[18+13\epsilon^2+40 \epsilon \cos{\psi}+9\epsilon^{2} \cos{2\psi}]\\ \nonumber
+&\frac{2G^{3}H^{2}}{15}~\frac{\mu^2 M^2 }{(1-\epsilon^2)^2~R^2}~
\big[60+100\epsilon^2+36\epsilon^4+\epsilon(180+113\epsilon^2)\cos\psi
+116\epsilon^2\cos2\psi+19\epsilon^3\cos3\psi\big]\\ \nonumber
-&\frac{32 G^{5/2}H^3}{5}~\frac{\epsilon^{2}\mu^2 M^{3/2}}{\sqrt{R(1-\epsilon^{2})}}~
\frac{\sin\psi(\epsilon+\cos\psi)}{1+\epsilon\cos\psi}\\ \nonumber
+&\frac{4G^{2}H^4}{15}~\frac{\mu^2 M R(1-\epsilon^2)}{(1+\epsilon\cos\psi)^{2}}~
\big[6+7\epsilon^{2}+12\epsilon\cos\psi {-}\epsilon^{2}\cos2\psi\big]
\end{align}
In the intermediate step we have  neglected pressure quadrupole moment term in eq. \eqref{RadiationField} and used $
\mathcal{Q}_{ij}\approx \partial_{t}^{3}Q_{ij}-3H\partial_{t}^{2}Q_{ij}+2H^{2}\partial_{t}
Q_{ij}$. To get the expression for power we have to perform an averaging integral of 
\eqref{InstPow}.
An explicit averaging procedure is illustrated in the appendix of \cite{DateJH2}. It permits to 
split four-dimensional 
averaging integral into an integral over $ r_{ph}=constant$ hypersurface and a three dimensional 
flux 
integral. The averaging integral over  $ r_{ph}=constant$ hypersurface and angular integral can be 
done 
explicitly, leaving the four-dimensional averaging integral to a time-averaged quantity only.
Hence average power over an orbital period is given by
\begin{align}
\mathcal{P}=&\frac{1}{T}\int_{0} ^{T} dt P(\psi)
=\frac{\omega_{s}}{2\pi} \int_{0} ^{2\pi} \frac{d\psi}{\dot{\psi}} P(\psi)\\
=&\bigg(\frac{GM}{R^{3}}\bigg)^{1/2}\times\frac{1}{2\pi}\int_{0} ^{2\pi} \frac{d\psi}
{\dot{\psi}} P(\psi)
\end{align}
Using \ref{InstPow} we obtain,
\begin{align}
\mathcal{P}=\frac{32G^{4}\mu^{2}M^3}{5 R^5}~ f(\epsilon)
+~\frac{8 H ^{2}G^{3} \mu^2 M^2}{ R^2} ~g(\epsilon)
+~ \frac{8 {H^{4}G^{2} \mu^{2} M R}}{5} ~h(\epsilon)
\end{align}
where,
\begin{align} \label{EnhanceFun1}
f(\epsilon)&=(1-\epsilon^2)^{-7/2}~\bigg(1+\frac{73}{24}\epsilon^{2}+\frac{37}{96}
\epsilon^{4}\bigg),\\
g(\epsilon)&=(1-\epsilon^{2})^{-1/2}~\bigg[\frac{4-\sqrt{1-\epsilon^ 2}}{3}\bigg],\\
h(\epsilon)&=1-\frac{\epsilon^2}{3}.  \label{EnhanceFun2}
\end{align}
Deriving this expression we have used $\dot{\psi}=\frac{\sqrt{GMR(1-
\epsilon^{2})}}{r_{ph}^{2}}$ for elliptic Keplerian orbit.
\\
\hspace*{0.5cm}
{We follow the same procedure for angular momentum flux. Neglecting pressure quadrupole 
moments and writing $\mathcal{L}_{T}$ in terms of $\partial_t$ the equation 
\eqref{AvgAngMom} becomes,
\begin{align}
\Big\langle\dot L_{z}\Big\rangle\approx \frac{2G}{5} \ \epsilon_{iz} \  ^{m} \Big\langle 
\big(\partial_{t}^{3}Q_{ij}-3H\partial_{t}^{2}Q_{ij}+2H^{2}\partial_{t} Q_{ij}\big) 
\big(\partial_{t}^{2}Q_{ij}-3H\partial_{t}Q_{ij}+2H^{2} Q_{ij}\big) \Big\rangle
\end{align}
 Plugging the derivatives of moments, the unaveraged quantity becomes,
 \begin{align}\label{InstAngPow}
\dot L_z\approx &\frac{4G^{7/2}}{5}~\frac{\mu^{2}M^{5/2}}{R^{7/2}(1-\epsilon^2)^{7/2}}
(1+\epsilon\cos{\psi})^{3}~\big[8+\epsilon^2+12 \epsilon \cos \psi+3\epsilon^2 \cos 2\psi
\big]\\ 
\nonumber
+&\frac{36G^{3}H}{5}~\frac{\epsilon~ \mu^{2}M^{2}}{R^{2}(1-\epsilon^{2})^{2}}~
\sin{\psi}(1+\epsilon \cos{\psi})^{2}~
%
\\ \nonumber
+&\frac{4G^{5/2}H^{2}}{5}~\frac{\mu^2 M^{3/2} }{(1-\epsilon^2)^{1/2}~R^{1/2}}~
\big[10+7\epsilon^2+17 \epsilon \cos\psi\big]\\ \nonumber
-&\frac{24 G^{2}H^3}{5}~
\frac{\epsilon\mu^2 M R\sin\psi}{1+\epsilon\cos\psi} \nonumber
+\frac{8G^{3/2}H^4}{5}~\frac{\mu^2 M^{1/2} R^{5/2}(1-\epsilon^2)^{5/2}}{(1+\epsilon\cos\psi)^{2}}~
\end{align}
Therefore, average angular momentum radiation over an orbital time period is given by,
\begin{align}
\Big\langle \dot L_{z}\Big\rangle&=\frac{1}{T}\int_{0} ^{T} d\tau ~ \frac{dL_{z}}
{d\tau} \\
& \approx  \frac{32G^{7/2}\mu^{2}M^{5/2}}{5  R^{7/2}} ~F(\epsilon)
+~\frac{8 H^{2} G^{5/2} \mu^2 M^{3/2}}{ R^{1/2}} ~G(\epsilon)
+~ \frac{8 H^{4} G^{3/2} \mu^{2} M^{1/2} R^{5/2}}{  5} ~ H(\epsilon)
\end{align}
where $F(\epsilon), G(\epsilon), H(\epsilon)$ are

\begin{align}
F(\epsilon)&=\frac{1}{(1-\epsilon^2)^{2}}~\big(1+\frac{7}{8}\epsilon^{2}\big),\\
G(\epsilon)&=(1-\epsilon^{2})^{1/2},\\
H(\epsilon)&=(1-\epsilon^{2})^{1/2}(1+\frac{3}{2}\epsilon^2).
\end{align}
\section{Weak field limit on de Sitter background} \label{Apn2}
 We follow the same procedure of Minkowski background to extract the 
 effective potential in de Sitter background. The equation for geodesic is 
 given by,
 \begin{equation}
 \frac{d^{2}x^\mu}{d\tau^{2}}+\Gamma^{\mu} _{\nu\rho} \frac{dx^{\nu}}
 {d\tau} \frac{dx^{\rho}}{d\tau}=0
 \end{equation}
 For our case the background de Sitter metric is given by,
 \begin{align}\label{background_g}
 \bar{ds}^{2}=-dt^{2}+e^{2Ht}dx^{i}dx^{j}\delta_{ij}
 \end{align}
 Hence, proper time $\tau$ is given by $t$ and for non-relativistic case, we 
 can neglect  $dx^{i}/dt$ terms. Hence the geodesic equation becomes,
 \begin{equation} \label{Geod}
 \frac{d^{2}x^i}{dt^{2}}+\Gamma^{i} _{00}=0
 \end{equation}
 Now decomposing the metric $g_{\mu\nu}=\bar{g}_{\mu\nu}+h_{\mu\nu}
 $, to the leading order in $h$, $\Gamma^{i}_{00}$ can be expressed as,
 \begin{equation}
 \Gamma^{i}_{00}=\bar{\Gamma}^{i}_{00}+\frac{1}{2}\bar{g}^{i\alpha}(2\bar{\nabla}_0 h_{\alpha 0}-\bar{\nabla}_{\alpha}h_{00})
 \end{equation}
 For background metric the only non-zero components of christoffel-connection are $\bar{\Gamma}
 ^{t}_{ij}=He^{2Ht}\delta_{ij}$ and $\bar{\Gamma}^{i}_{tj}=H\delta^{i}_{j}$. Therefore,
 \begin{equation}
 2\bar{\nabla}_0 h_{j 0}-\bar{\nabla}_{j}h_{00}=\partial_{0}h_{j0}-\partial_{j}h_{00}
 \end{equation}
 Since we are considering non-relativistic source, the time derivative of metric is of higher order 
 with respect to spatial derivatives ($\frac{\partial}{\partial t}=\mathcal{O}(v)\frac{\partial}{\partial x}$), so to the leading order eqn. 
 \eqref{Geod} becomes,
 \begin{equation}
 \frac{d^{2}x^{i}}{dt^{2}}=\frac{1}{2} \bar{g}^{ij} \partial_{j}h_{00}
 \implies A^{i}=\frac{1}{2} e^{-2Ht} \delta^{ij} \partial_{j} h_{00}
 \end{equation}
 As the orbital parameters are described in tetrad frame, we would like to recast this expression 
 in that frame. For the background metric \eqref{background_g}, tetrad frame becomes ,
$ f^{\alpha}_{\underline{0}}=\delta^{\alpha}_0, f^{\alpha}_{\underline{i}}=e^{-Ht}\delta^{\alpha}
 _{\underline{i}}$. Therefore $A^{i}=e^{-Ht}\bar{A}^{i}$ and $\frac{\partial}{\partial x^{j}}=e^{Ht}\frac{\partial}{\partial\bar{x}^j}$
 Therefore in this frame geodesic equation becomes,
 \begin{equation}
 \bar{A}^{i}=\frac{1}{2}\delta^{ij} \frac{\partial h_{00}}{\partial\bar{x}^{j}}
 \end{equation}
 Comparing this equation with Newtonian equation of motion, the term $-\frac{1}{2}h_{00}$ can be thought of as Newtonian potential.
 Under the coordinate transformations, 
 \begin{equation}
 dt=d\tau-\frac{Hr_{ph}}{1-H^2 r_{ph}^{2}}dr_{ph}~, ~~r=e^{-Ht} r_{ph}
 \end{equation}
 background metric \eqref{background_g} becomes,
 \begin{equation}
 \bar{ds}^{2}=-(1-H^2 r_{ph}^{2})d\tau^{2}+\frac{dr_{ph}^{2}}{(1-H^2 
 r_{ph}^{2})}+r_{ph}^{2}(d\theta^{2}+\sin^{2}\theta d\phi^{2})
 \end{equation}
 It should be noted that tetrad frame measure physical distance, $r_{ph}$ and time direction is 
 taken along worldline of centre of mass (for which $r_{ph}=0$). Therefore $dt=d\tau$ in this 
 frame.
Now in Schwarschild de Sitter geometry the metric takes the form,
\begin{equation} \label{SdS}
ds^{2}=-(1-H^2 r_{ph}^{2}-\frac{2Gm}{r_{ph}})d\tau^{2}+\frac{dr_{ph}
^{2}}{(1-H^2 
 r_{ph}^{2}-\frac{2Gm}{r_{ph}})}+r_{ph}^{2}(d\theta^{2}+\sin^{2}\theta 
 d\phi^{2})
\end{equation}
Therefore, on pure de Sitter background the motion is still governed by 
Newtonian potential in weak field limit.
 }


\begin{thebibliography}{10}
%
%
%
\bibitem{Einstein} A. Einstein, \"{ U}ber Gravitationswellen, Sitzungsberichte der K\"{o}niglich 
Preu{\ss}ischen Akademie der Wissenschaften (Berlin), 154-167  (1918).
%
\bibitem{TaylorI} R.A. Hulse and J.H. Taylor, Discovery of a pulsar in a binary 
system, {\em Astrophys. J.}, {\bf 195}, L51-L53 (1975).
%
\bibitem{TaylorII} J.H. Taylor and J. M. Weisberg, A new test of general 
relativity :  Gravitational radiation and the binary PSR 1913+1916, {\em Astrophys. J.}, {\bf 253}, 908-920 (1982).
%
\bibitem{Damour} T. Damour, 1974 : The discovery of the first binary pulsar, 
{\em Class. Quant. Grav.}, {\bf 32}, 124009 (2015).
%
\bibitem{GWI}  B. P. Abbott et al., Observation of Gravitational Waves from a Binary Black Hole 
Merger, {\em Phys. Rev. Lett.}, {\bf 116}, 061102 (2016).
%
\bibitem{GWII} B. P. Abbott et al.,  GW151226: Observation of Gravitational Waves from a 22-
Solar-Mass Binary Black Hole Coalescence, {\em Phys. Rev. Lett.}, {\bf 116}, 241103  (2016).
%
\bibitem{GWIII} B. P. Abbott et al., GW170104: Observation of a 50-Solar-Mass Binary Black 
Hole  Coalescence at Redshift 0.2, {\em Phys. Rev. Lett.}, {\bf 118}, 221101  (2017).
%
\bibitem{GWIV} B. P. Abott et al., GW170814 : A three-detector observation of gravitational waves from a binary black hole coalescence, [arXiv:1709.09660].
%
\bibitem{Ankit} A. Aggarwal, {  Estimating the effects of a small positive $\Lambda$ in orbital 
decay of binaries}, Master's thesis, June 2017.
\bibitem{Jahanur} S. J. Hoque, { Physics of Gravitational Waves in 
Presence of Positive Cosmological Constant},  PhD thesis, June 2017.
%
\bibitem{Isaacson} R. A. Isaacson, Gravitational radiation in the limit
of high frequency I: The linear approximation and geometrical optics,
{\em Phys. Rev.}, {\bf 166}, 1263, (1968);

R. A. Isaacson, Gravitational radiation in the limit of high frequency
II: Non-linear terms and the effective stress tensor, {\em Phys. Rev.},
{\bf 166}, 1272, (1968).
%
\bibitem{DateJH2} G. Date and S. J. Hoque, Cosmological Horizon and the 
Quadrupole Formula in de Sitter Background, {\em Phys. Rev. D}, {\bf 96}, 
044026 (2017).
%
\bibitem{DateJHI} G. Date and  S. J. Hoque, Gravitational Waves from 
Compact Sources in de Sitter 
Background, {\em Phys. Rev. D}, { \bf 94}, 064039 (2016).
%
\bibitem{ABKII} A. Ashtekar, B. Bonga and A. Kesavan,
Asymptotics with a positive cosmological constant: II. Linear fields on
de Sitter space-time, {\em Phys. Rev. D}, {\bf 92}, 044011, (2015),
[arXiv:1506.06152]. 
%
\bibitem{ABKIII}A. Ashtekar, B. Bonga and A. Kesavan, Asymptotics with
a positive cosmological constant: III. The quadrupole formula, {\em
Phys. Rev. D}, {\bf 92}, 10432 (2015), [arXiv:1510.05593]\ .
%
\bibitem{JHIII} S. J. Hoque, A. Virmani, On Propagation of Energy Flux in de Sitter Spacetime, {\em Gen. Rel. Grav. }, {\bf 50}, 4, 40 (2018), [arXiv: 1801.05640].
\bibitem{MTW} Charles W. Misner, Kip S. Thorne, and John Archibald Wheeler, {\em Gravitation}, 
Macmillan, (1973).
%
\bibitem{Stein} Leo C. Stein and N. Yunes, Effective Gravitational Wave Stress-energy Tensor in 
Alternative Theories of Gravity, {\em Phys. Rev. D}, {\bf 83}, 064038, (2011), [arXiv:
1012.3144].
%
\bibitem{Gravity} E. Poisson and C.M.Will, Gravity : Newtonian, Post-Newtonian, Relativistic, CUP, 
2014.
%


\bibitem{Poisson} E.Poisson, A.Pound and l. Vega, The motion of point particles in curved space-
time, {\em Living Rev. Relativity}, {\bf 14}, 7, (2011), [arXiv:1102.0529].
%
\bibitem{Straumann} N. Straumann, { General Relativity with Applications to Astrophysics }, 
Springer, Berlin, 2004.
%
\bibitem{Maggiore} M. Maggiore, { Gravitational waves}, Vol 1: Theory and experiments, OUP, 2007.
%
\bibitem{Bounanno} A. Bounanno, Gravitational waves, [arXiv: 0709.4682].
%
\bibitem{Peters} P.C. Peters, J. Mathews, Gravitational radiation from point masses in a
Keplerian orbit , {\em Phys. Rev. }, {\bf 131},  435-440 (1963).
%
%
%
\bibitem{Satya} B.S. Sathyaprakash, B.F. Schutz, Physics, Astrophysics and 
Cosmology with Gravitational Waves, {\em Living Rev. Relativity}, {\bf 12}, 
(2009),  [arXiv: 0903.0338].
%
\bibitem{Bonga} B. Bonga, J. S. Hazboun, Power radiated by a binary system in a de Sitter Universe,  {\em Phys. Rev. D}, {\bf 96}, 064018, (2017), [arXiv:1708.05621].
%
\bibitem{ABI} A. Ashtekar, B. Bonga, On a basic conceptual confusion in gravitational radiation 
theory, {\em Class. Quantum Grav.}, { \bf 34}, 20LT01,  (2017),  [arXiv:1707.07729].
%
\bibitem{ABII}  A. Ashtekar, B. Bonga, On the ambiguity in the notion of transverse traceless 
modes of gravitational waves, { \em Gen. Rel. Gravit.}, {\bf 49}, 122,  (2017),  [arXiv:1707.09914].
%
%
\end{thebibliography}
\end{document}